# 基於偽隨機數產生器一次性簽章

Abel C. H. Chen*

Information & Communications Security Laboratory,
Chunghwa Telecom Laboratories

**摘要**

有鑑於量子計算的技術日益成熟，所以近幾年陸續盤點具備抗量子計算攻擊的密碼學方法，並且建立後量子密碼學方法。其中，基於雜湊(Hash-based)數位簽章演算法(Digital Signature Algorithm, DSA)是後量子密碼學方法之一，並且雜湊計算除了應用在數位簽章外，也常應用到偽隨機數產生器(Pseudorandom Number Generator, PRNG)。因此，本研究主要在基於雜湊數位簽章演算法的基礎上修改，提出基於偽隨機數產生器數位簽章演算法，可以適用在一次性簽章(One-Time Signature, OTS)的應用情境。本研究中討論提出的基於偽隨機數產生器一次性簽章的安全性，並且通過實驗比較不同參數組合下的金鑰長度、簽章長度、金鑰產製時間、簽章產製時間、以及簽章驗章時間，以證明提出方法的可行性。

# One-Time Signature Based on Pseudorandom Number Generator

**Abstract**

With the advancement of quantum computing technologies, recent years have seen increasing efforts to identify cryptographic methods resistant to quantum attacks and to establish post-quantum cryptography (PQC) approaches. Among these, hash-based digital signature algorithms (DSAs) are a notable category of PQC. Hash functions are not only utilized in digital signatures but are also widely applied in pseudorandom number generators (PRNGs). Building on the foundation of hash-based DSAs, this study proposes a modified approach that introduces a DSA based on PRNGs, suitable for one-time signature (OTS) applications. The study explores the security of the proposed PRNG-based OTS algorithm and validates its feasibility through experiments comparing various parameter configurations. These experiments examine key length, signature length, key generation time, signature generation time, and signature verification time under different parameter settings.

## 1.前言

近年來為提升抵抗量子計算攻擊的能力，後量子密碼學開始蓬勃發展，美國

* Corresponding author E-mail: chchen.scholar@gmail.com

國家標準暨技術研究院(National Institute of Standards and Technology, NIST)[1]整理出基於雜湊(Hash-based)密碼學方法[2]、基於晶格(Lattice-based)密碼學方法[3]、基於編碼(Code-based)密碼學方法[4]、以及基於多變量(Multivariate-based)密碼學方法[5]等。其中，美國國家標準暨技術研究院在 2024 年 8 月更確定了後量子密碼學最終標準，包含無狀態雜湊數位簽章演算法(Stateless Hash-based Digital Signature Algorithm, SLH-DSA)[6]、模晶格數位簽章演算法(Module-Lattice-based Digital Signature Algorithm, ML-DSA)[7]、以及模晶格金鑰封裝機制(Module-Lattice-based Key-Encapsulation Mechanism, ML-KEM)[8]等。由此可知，基於雜湊數位簽章演算法仍是抵抗量子計算攻擊的重要方法之一。

在基於雜湊數位簽章演算法系列中，Winternitz 一次性簽章(One-Time Signature, OTS)演算法[9]是經典方法之一。除此之外，雜湊計算除了應用在數位簽章外，也常用在偽隨機數產生器[10]。有鑑於此，本研究主要在 Winternitz 一次性簽章基礎上，融入偽隨機數產生器，提出基於偽隨機數產生器數位簽章演算法，可以適用在一次性簽章的應用情境。本研究的主要貢獻條列如下：

- 本研究提出基於偽隨機數產生器一次性簽章演算法，可以建構在 Winternitz 一次性簽章和線性同餘產生器(Linear Congruential Generator, LCG)[11]的基礎上來提供另一種數位簽章演算法的候選方法。
- 本研究提供數學理論證明基於偽隨機數產生器一次性簽章演算法的可行性，並且提供實例展示計算過程。
- 本研究實作各種線性同餘產生器的參數組合，包含常見的 Microsoft Visual Basic 採用的偽隨機數產生器參數[12]、GNU Compiler Collection (GCC)採用的偽隨機數產生器參數[13]、POSIX (Portable Operating System Interface)採用的偽隨機數產生器參數[14]、MMIX 採用的偽隨機數產生器參數[15]。本研究比較和驗證在不同參數組合下的金鑰長度、簽章長度、金鑰產製時間、簽章產製時間、以及簽章驗章時間。

本論文總共包含五個章節。第 2 節主要介紹基於線性同餘產生器的偽隨機數產生器和基於雜湊一次性簽章演算法。第 3 節詳細說明本研究提出的基於偽隨機數產生器一次性簽章演算法，並且提供理論證明和計算實例。第 4 節描述實驗環境和實驗結果與討論。最後，第 5 節總結本研究貢獻和討論未來發展方向。

# 2.文獻探討

在第 2.1 節將先介紹基於線性同餘產生器的偽隨機數產生器，說明線性同餘產生器的原理，再描述如何運用線性同餘產生器產生偽隨機數。在第 2.2 節將介紹基於雜湊一次性簽章演算法，並以 Winternitz 一次性簽章演算法為例說明。

## 2.1 基於線性同餘產生器的偽隨機數產生器

本節將先說明基於線性同餘產生器的偽隨機數產生器的演算法設計，再舉計算實例展示。

### 2.1.1 演算法設計

為更明確說明基於線性同餘產生器的偽隨機數產生器，本節以 Java 預設的偽隨機數產生器(也就是與 POSIX 採用的偽隨機數產生器參數一樣)為例說明。

首先，可以讓使用者輸入一隨機數種子(seed)整數 $s$，並且根據公式(1)產生初始值$f_0(s)$。需要注意的是不同的偽隨機數產生器其產生初始值的作法可能不同，在 Java 預設的偽隨機數產生器是用隨機數種子與乘數 $a$ 做邏輯互斥或(Exclusive-OR) $\oplus$ 計算，再模整數 $m$ 以限制隨機數長度。

$$f_0(s) = a \oplus s \ (\text{mod } m) \tag{1}$$

在產生隨機數的過程主要採用線性同餘產生器，其計算方式如公式(2)所示。當產生第 $n$ 個隨機數$f_n(s)$時，主要計算乘數 $a$ 和第 $n-1$ 個隨機數$f_{n-1}(s)$相乘後再加上一增量整數 $c$，再模整數 $m$ 限制隨機數長度，以遞迴的方式來產生。

$$f_n(s) = a \times f_{n-1}(s) + c \ (\text{mod } m) \tag{2}$$

### 2.1.2 計算實例

本節提供 Java 預設的偽隨機數產生器實例說明；其中，線性同餘產生器參數值包含如下：$a = 25214903917$、$c = 11$、$m = 2^{48}$，以及本例設定隨機數種子$s = 1$。以下說明過程皆以 16 進位 Hex 值表示，運用公式(1)，代入隨機數種子$s = $ 0x01 後可以得到初始值$f_0(\text{0x01}) = $ 0x05DEECE66C，如公式(3)所示。

$$
\begin{aligned}
f_0(0x01) &= 0x05DEECE66D \oplus 0x01 \pmod{0x01000000000000} \\
&= 0x05DEECE66C \pmod{0x01000000000000} \\
&= 0x05DEECE66C
\end{aligned} \tag{3}
$$

產生隨機數時，可以以遞迴方式來執行公式(2)；例如，產生第 1 個隨機數$f_1(0x01)$的計算方式如公式(4)所示，以依產生第 2 個隨機數$f_2(0x01)$的計算方式如公式(5)所示。依此類推，可產生後續的隨機數。

$$
\begin{aligned}
f_1(0x01) &= 0x05DEECE66D \times f_0(0x01) + 0x0B \\
&\qquad \pmod{0x01000000000000} \\
&= 0x227760BB1AD57323FC + 0x0B \\
&\qquad \pmod{0x01000000000000} \\
&= 0x227760BB1AD5732407 \pmod{0x01000000000000} \\
&= 0xBB1AD5732407
\end{aligned} \tag{4}
$$

$$
\begin{aligned}
f_2(0x01) &= 0x05DEECE66D \times f_1(0x01) + 0x0B \\
&\qquad \pmod{0x01000000000000} \\
&= 0x044A74958019B89CD8A0FB + 0x0B \\
&\qquad \pmod{0x01000000000000} \\
&= 0x044A74958019B89CD8A106 \pmod{0x01000000000000} \\
&= 0x19B89CD8A106
\end{aligned} \tag{5}
$$

## 2.2 基於雜湊一次性簽章演算法

本節將先說明 Winternitz 一次性簽章演算法的演算法設計，再舉計算實例展示。

### 2.2.1 演算法設計

在 Winternitz 一次性簽章演算法產製金鑰對時，將先產生隨機數整數值$r$作為私鑰值，再對該整數做$2^w - 1$次雜湊，取雜湊後的值作為公鑰值$R$，如公式(6)所示；其中，$h(\cdot)$表示雜湊函數，$h_n(r)$表示對$r$做$n$次雜湊計算，$h_n(r)$和$h_{n-1}(r)$關係如公式(7)所示。

$$R = h_{2^w-1}(r) \tag{6}$$

$$h_n(r) = h\big(h_{n-1}(r)\big) = h_{n-1}\big(h(r)\big) \tag{7}$$

在產製簽章時，首先把待簽章訊息正規化為$[0, 2^w - 1]$區間，常用的正規化方法是對訊息做雜湊計算，並且取最後$w$個位元值。假設待簽章訊息正規化後的值為$t$，由簽章方為對私鑰值$r$做$t$次雜湊，取雜湊後的值作為簽章值$\zeta$，如公式(8)所示。在驗證簽章時，驗證方可以取得簽章方的公鑰值$R$、待簽章訊息正規化後的值$t$、以及簽章值$\zeta$，再對簽章值$\zeta$做$2^w - 1 - t$次雜湊，取雜湊後的值作為驗證值$\xi$，如公式(9)所示；再比對公鑰值$R$和驗證值$\xi$是否一致，如果一致表示驗章通過，如果不一致則表示驗章未通過。

$$\zeta = h_t(r) \tag{8}$$

$$\begin{aligned} \xi &= h_{2^w-1-t}(\zeta) \\ &= h_{2^w-1-t}\big(h_t(r)\big) = h_{2^w-1}(r) = R \end{aligned} \tag{9}$$

### 2.2.2 計算實例

為說明 Winternitz 一次性簽章演算法，以下提供計算實例說明，並為了簡短說明，本節採用 SHA-224 作為雜湊函數。

本例的私鑰值 $r$ 採用隨機生的整數值 0xD14A028C2A3A2BC9476102BB288234C415A2B01F828EA62AC5B3E42F，並且設定 $w$ 為 24，對私鑰值做 16777215 次雜湊計算，取得雜湊計算結果 0xF9DAF6920241798166A3D933188EB066126F0F791394AD27F1B3024A 作為公鑰值$R$，如公式(10)和公式(11)所示。

$$r = \text{0xD14A028C2A3A2BC9476102BB288234C415A2B01F828EA62AC5B3E42F} \tag{10}$$

$$R = h_{16777215}(r) = \text{0xF9DAF6920241798166A3D933188EB066126F0F791394AD27F1B3024A} \tag{11}$$

在產製簽章時，假設待簽章訊息正規化的十進位值為 12345678，對私鑰值$r$做 12345678 次雜湊計算得簽章值 0x30A2839E846E948517123CEFC4A32DDB42AEA6CED1FD81D1DDC8E4F2，如公式(12)所示。在驗證簽章時，對簽章值$\zeta$做$2^w - 1 - t = 16777215 - 12345678 = 4431537$次雜湊，取得雜湊結果為驗章值 0xF9DAF6920241798166A3D933188EB066126F0F791394AD27F1B3024A，如公式(13)所示。比對驗章值和公鑰值，在此例中比對結果一致，所以驗章通過。

$$\zeta = h_{12345678}(r) = \text{0x30A2839E846E948517123CEFC4A32DDB42AEA6CED1FD81D1DDC8E4F2} \tag{12}$$

$$\begin{aligned} \xi &= h_{4431537}(\zeta) \\ &= h_{4431537}\big(h_{12345678}(r)\big) \\ &= h_{16777215}(r) = \\ &\text{0xF9DAF6920241798166A3D933188EB066126F0F791394AD27F1B3024A} \end{aligned} \tag{13}$$

# 3.本研究提出的基於偽隨機數產生器一次性簽章

在第 3.1 節先介紹本研究提出的基於偽隨機數產生器一次性簽章演算法的設計。在第 3.2 節再提供計算實例展示本研究提出的基於偽隨機數產生器一次性簽章演算法。最後，在第 3.3 節對本研究提出的基於偽隨機數產生器一次性簽章演算法的安全性進入深入討論。

## 3.1 設計理念

本節將對本研究提出的基於偽隨機數產生器一次性簽章演算法分別從金鑰產製、簽章產製、以及簽章驗證各別說明。最後，再提供數學理論證明本研究提出的基於偽隨機數產生器一次性簽章演算法。

### 3.1.1 金鑰產製

本研究提出的基於偽隨機數產生器一次性簽章演算法在產製金鑰對時，首先將產製產生隨機數種子值$p$作為私鑰值，通過公式(1)計算初始值後，再運用公式(2)取得$2^w-1$個隨機數整數值作為公鑰值$P$，如公式(14)和公式(15)所示；其中，$f_{2^w-1}(p)$表示線性同餘產生器，對$f_0(p)$做$2^w-1$次線性同餘方法計算。

$$f_0(p) = a \oplus p \pmod{m} \tag{14}$$

$$P = f_{2^w-1}(p) = a \times f_{2^w-2}(p) + c \pmod{m} \tag{15}$$

### 3.1.2 簽章產製

在產製簽章時，首先把待簽章訊息正規化為$t$，並且$t$值介於$[0, 2^w-1]$區間，再由簽章方為對私鑰值$p$作為隨機數種子以公式(16)取得第$t$個隨機數整數值與乘數 $a$ 做邏輯互斥或(Exclusive-OR) $\oplus$ 計算，再模整數 $m$ 後作為簽章值$S$，如公式(17)所示。

$$f_t(p) = a \times f_{t-1}(p) + c \pmod{m} \tag{16}$$

$$S = a \oplus f_t(p) \pmod{m} \tag{17}$$

### 3.1.3 簽章驗證

在驗證簽章時，驗證方將先取得簽章方的公鑰值$P$、待簽章訊息正規化後的值$t$、以及簽章值$S$，再把簽章值$S$作為隨機數種子後，取得第$2^w-1-t$個隨機數作為驗證值$V$，如公式(18) 和公式(19)所示；再比對公鑰值$P$和驗證值$V$是否一致，如果一致表示驗章通過，如果不一致則表示驗章未通過。

$$f_0(S) = a \oplus S \ (\mathrm{mod}\ m) \tag{18}$$

$$\begin{aligned} V &= f_{2^w-1-t}(S) \\ &= a \times f_{2^w-2-t}(S) + c \ (\mathrm{mod}\ m) \end{aligned} \tag{19}$$

### 3.1.4 理論證明

在本節主要證明為什麼簽章值$S$作為隨機數種子後，取得第$2^w - 1 - t$個隨機數作為驗證值$V$，該驗證值$V$會等於公鑰值$P$。其中，當把簽章值$S$作為隨機數種子後，初始狀態會是，如公式(20)所示。因此，當取得第$2^w - 1 - t$個隨機數$V$作為驗證值時，該驗證值$V$會與公鑰值$P$一致，如公式(21)所示。

$$\begin{aligned} f_0(S) &= a \oplus S \ (\mathrm{mod}\ m) \\ &= a \oplus a \oplus f_t(p) \ (\mathrm{mod}\ m) \\ &= f_t(p) \ (\mathrm{mod}\ m) \end{aligned} \tag{20}$$

$$\begin{aligned} V &= f_{2^w-1-t}\big(f_t(p)\big) \\ &= f_{2^w-1}(p) = P \end{aligned} \tag{21}$$

## 3.2 計算實例

本節提供的計算實例主要採用 Java 預設的偽隨機數產生器(也就是與 POSIX 採用的偽隨機數產生器參數一樣)[14]：$a = 25214903917$、$c = 11$、$m = 2^{48}$。

### 3.2.1 金鑰產製

本計算實例採用的私鑰值$p$是 0x13579BDE，作為隨機數種子值，通過公式(1)計算初始值，如公式(22)所示。再運用線性同餘產生器，基於公式(2)取得$2^w - 1$個隨機數整數值作為公鑰值$P$，設定 $w$ 為 24，如公式(23)所示。

$$\begin{aligned} f_0(p) &= f_0(\mathrm{0x13579BDE}) \\ &= \mathrm{0x05DEECE66D} \oplus \mathrm{0x13579BDE} \ (\mathrm{mod}\ \mathrm{0x01000000000000}) \end{aligned} \tag{22}$$

$$f_{2^w-1}(p) = f_{16777215}(0\text{x}13579\text{BDE}) = 0\text{x}05\text{DEECE}66\text{D} \oplus f_{16777214}(0\text{x}13579\text{BDE}) + 0\text{x}0\text{B} \pmod{0\text{x}01000000000000} = 0\text{xE}9694\text{A}840\text{B}48 \quad (23)$$

### 3.2.2 簽章產製

在產製簽章時，假設待簽章訊息正規化的十進位值$t$為 12345678，由簽章方為對私鑰值$p$作為隨機數種子以公式(24)取得第 12345678 個隨機數整數值 0xECE653813E21。再把與乘數 $a$ 做邏輯互斥或(Exclusive-OR) ⊕ 計算，再模整數 $m$ 後作為簽章值$S$為 0xECE38D6DD84C，如公式(25)所示。

$$f_t(p) = f_{12345678}(0\text{x}13579\text{BDE}) = 0\text{x}05\text{DEECE}66\text{D} \times f_{12345677}(0\text{x}13579\text{BDE}) + 0\text{x}0\text{B} \pmod{0\text{x}01000000000000} = 0\text{xECE}653813\text{E}21 \quad (24)$$

$$S = a \oplus f_t(p) \pmod{m} = 0\text{x}05\text{DEECE}66\text{D} \oplus 0\text{xECE}653813\text{E}21 \pmod{0\text{x}01000000000000} = 0\text{xECE}38\text{D}6\text{DD}84\text{C} \quad (25)$$

### 3.2.3 簽章驗證

在驗證簽章時，把簽章值$S$作為隨機數種子後，取得第$2^w-1-t$個隨機數作為驗證值$V$，如公式(26) 和公式(27)所示。比對驗章值和公鑰值，在此例中比對結果一致，所以驗章通過。

$$f_0(S) = a \oplus S \pmod{m} = \; = 0\text{x}05\text{DEECE}66\text{D} \oplus 0\text{xECE}38\text{D}6\text{DD}84\text{C} \pmod{0\text{x}01000000000000} = 0\text{xECE}653813\text{E}21 \quad (26)$$

$$V = f_{2^w-1-t}(S)$$
$$= f_{4431537}(0\text{xECE38D6DD84C}) \quad (27)$$
$$= 0\text{xE9694A840B48}$$

## 3.3 安全性討論

本研究提出的基於偽隨機數產生器一次性簽章演算法主要建構在基於線性同餘產生器的偽隨機數產生器，並且線性同餘產生器計算方法如公式(2)所示。由公式可知由於第 $n$ 個隨機數$f_n(s)$是由第 $n-1$ 個隨機數$f_{n-1}(s)$經過乘法和加法運算後再做模數計算，所以想從第 $n$ 個隨機數$f_n(s)$回推第 $n-1$ 個隨機數$f_{n-1}(s)$時將存在多種可能性，並且無法得到唯一解。因此，在本研究提出的基於偽隨機數產生器一次性簽章演算法，攻擊者無法從公鑰回推私鑰。

在產製簽章時，由於攻擊者在不知道私鑰值的情況下，無法偽造簽章，所以如果在一次性簽章的情況下是具備安全性的。但如果同一金鑰對用多個簽章時，則可能存在被偽造簽章的風險，其限制與 Winternitz 一次性簽章演算法相似。

除此之外，在計算例提供的私鑰長度較短，建議在實際應用時可以採用較長的金鑰長度和線性同餘產生器的參數值，以提升安全性。

# 4.實驗結果與討論

本節將先介紹實驗環境，然後再針對各種線性同餘產生器參數組合進行實作和比較。

## 4.1 實驗環境

本研究主要採用 Java 內建的 BigInteger 類似來實作各種偽隨機數產生器，各個偽隨機數產生器的線性同餘產生器參數表如表 1 所示。每個偽隨機數產生器都是自行復刻和開發，並且在此基礎上建立本研究提出的基於偽隨機數產生器一次性簽章演算法，並且進行計算效率比較。本研究具體採用的硬體和軟體規格為：Intel(R) Core(TM) i7-10510U CPU、16 GB RAM、Windows 10 Enterprise、OpenJDK 18.0.2.1。

**表 1　各個偽隨機數產生器的線性同餘產生器參數表**

| 偽隨機數產生器 | *a* | *c* | *m* |
|---|---|---|---|
| [12] | 16598013 | 12820163 | $2^{24}$ |
| [13] | 1664525 | 1013904223 | $2^{31}$ |
| [14] | 25214903917 | 11 | $2^{48}$ |
| [15] | 6364136223846793005 | 1442695040888963407 | $2^{64}$ |

## 4.2 實驗結果

本研究主要比對各個偽隨機數產生器的線性同餘產生器參數組合下的私鑰長度、公鑰長度、以及簽章長度，如表 2 所示。其中，由於長度的部分主要受到模數 $m$ 限制，所以當模數為 $2^{24}$ 時，則私鑰長度、公鑰長度、以及簽章長度都是 24 位元。然而，雖然長度越短，使用的儲存空間越少且傳輸時的時間越短，但相對較不安全。當採用同一種線性同餘產生器方法計算時，模數越大，則相對越安全。

**表 2　私鑰長度、公鑰長度、以及簽章長度比較表(單位：位元)**

| 參數 | 私鑰長度 | 公鑰長度 | 簽章長度 |
|---|---|---|---|
| 本研究方法基於[12]的參數 | 24 | 24 | 24 |
| 本研究方法基於[13]的參數 | 31 | 31 | 31 |
| 本研究方法基於[14]的參數 | 48 | 48 | 48 |
| 本研究方法基於[15]的參數 | 64 | 64 | 64 |

金鑰產製時間、簽章產製時間、以及簽章驗證時間，如表 3 所示。其中，可以觀察到採用文獻[13]的參數組合時可以有最短的計算時間，不論是在金鑰產製時間、簽章產製時間、以及簽章驗證時間上都有最短計算時間的表現；其原因主要在於文獻[13]的參數組合中的 a 值最小，所以乘法計算後的值可以最小，相對可以有最短的計算時間，但也反應出可能存在安全上的風險。相反的，雖然採用文獻[14]和[15]的參數組合時都有最長的計算時間，不論是在金鑰產製時間、簽章產製時間、以及簽章驗證時間上都有最長計算時間的表現；其原因主要在於文獻[14]和[15]的參數組合中的各種參數值都是最大的，所以進行相關計算後的值也是最大，相對需要最長的計算時間，但也具備最高的安全性。

**表 3　金鑰產製時間 簽章產製時間、以及簽章驗證時間比較表(單位：毫秒)**

| 參數 | 金鑰產製時間 | 簽章產製時間 | 簽章驗證時間 |
|---|---|---|---|
| 本研究方法基於[12]的參數 | 926.945 | 373.617 | 568.218 |
| 本研究方法基於[13]的參數 | 800.579 | 347.363 | 460.379 |
| 本研究方法基於[14]的參數 | 2912.141 | 1835.912 | 1081.136 |
| 本研究方法基於[15]的參數 | 2699.229 | 1112.627 | 1588.626 |

為更清楚展示計算時間的分佈，本研究採用盒鬚圖顯示金鑰產製時間、簽章產製時間、以及簽章驗證時間，如圖 1、圖 2、圖 3 所示。從分佈可以清楚觀察到，採用文獻[13]的參數組合時有顯著較短的計算時間，而採用文獻[14]和[15]的參數組合時則有顯著較長的計算時間，如前段所述一致。

# 5.結論與未來研究

本研究提出原創的基於偽隨機數產生器一次性簽章演算法，可以建構在線性同餘產生器的基礎上來產生，並且確保其安全性，在第 3 節提供詳細的數學理論證明·此外，本研究在第 4 節提供在真實環境下的計算效率比較，提供實證分析。

在未來研究中，可以考慮設計更長的線性同餘產生器來提升安全性。此外，可以考慮在線性同餘產生器計算上加速，以提供快速且簡短的簽章效果。

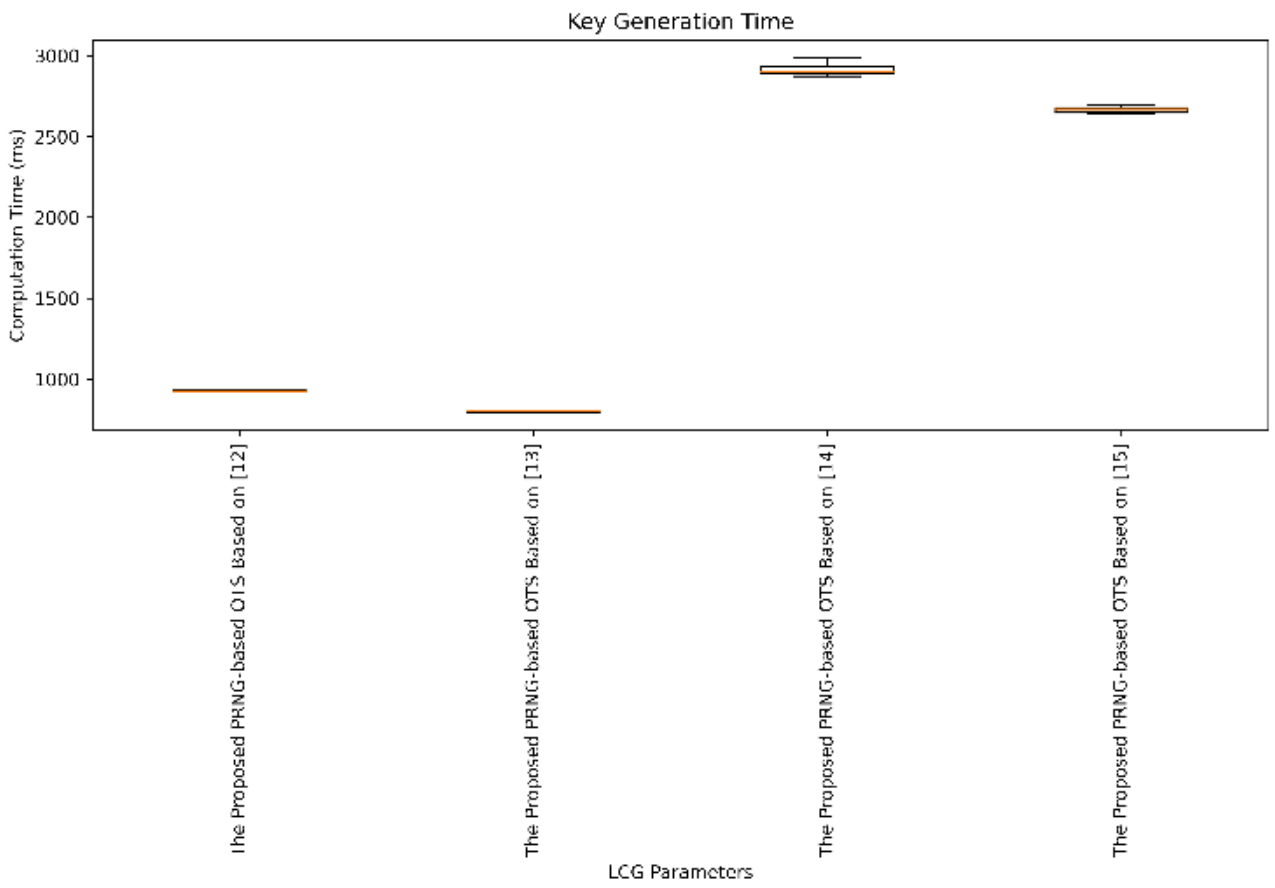

**圖 1　金鑰產製時間比較**

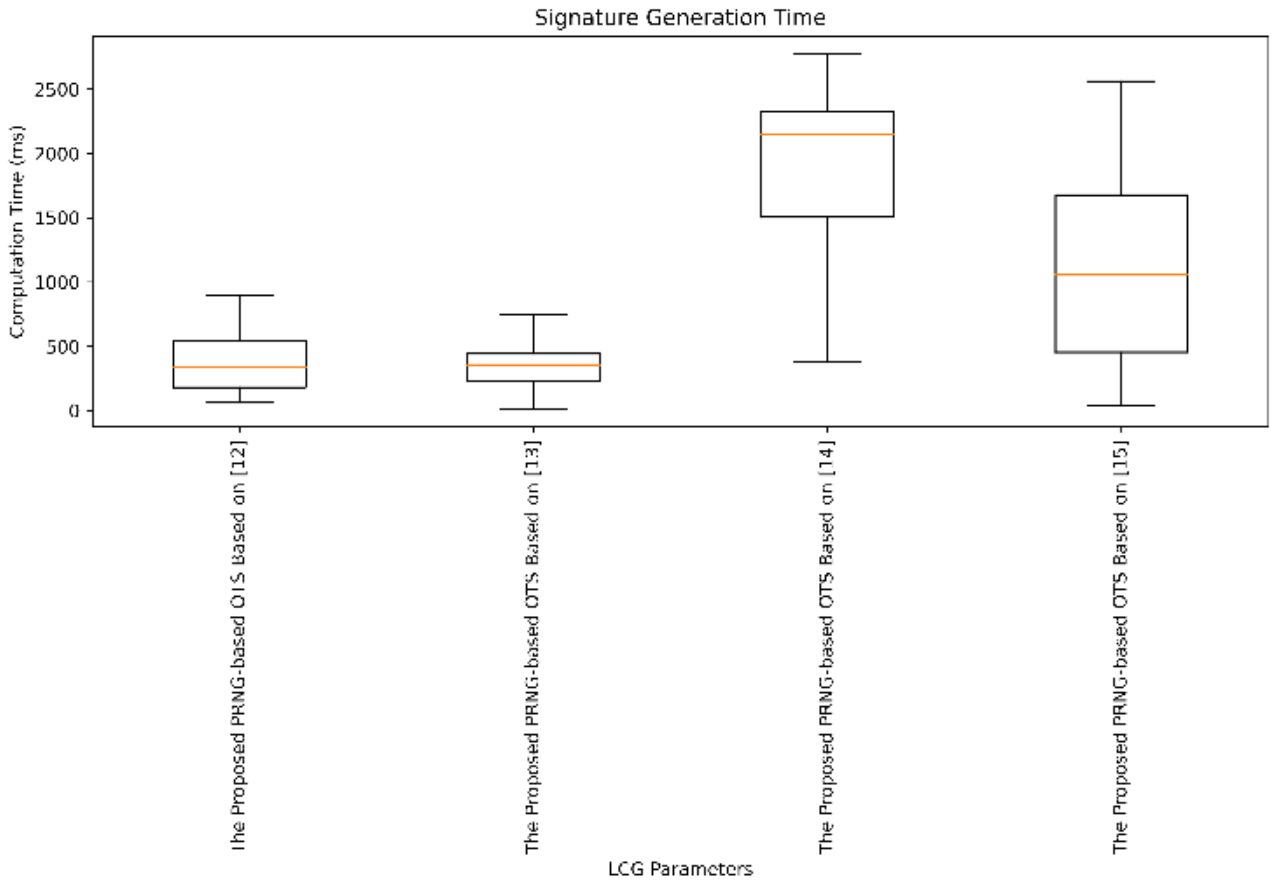

**圖 2　簽章產製時間比較**

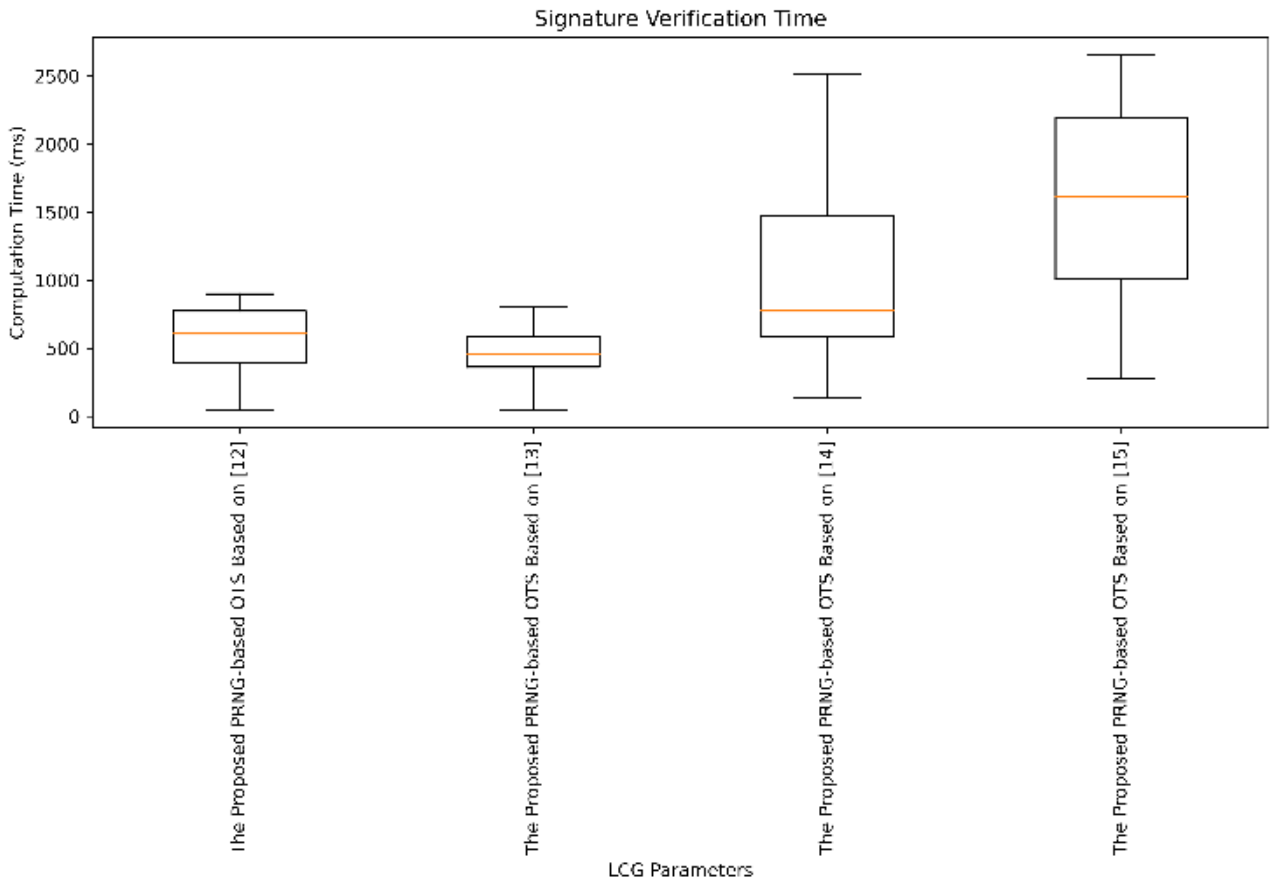

**圖 3　簽章驗證時間比較**

# 參考文獻